\documentclass[final,authoryear,5p,times,twocolumn]{elsarticle}
\usepackage{graphicx,longtable}
\usepackage{natbib}
\usepackage{amssymb}
\usepackage{amsthm}
\journal{New Astronomy}
\newcommand{\msun}{\mbox{\rm $M_{\odot}$~}}

\newcommand{\hii}{\mbox{H{\sc ii}~}}

\def\gsim{\;\rlap{\lower 2.5pt\hbox{$\sim$}}\raise 1.5pt\hbox{$>$}\;}
\def\lsim{\;\rlap{\lower 2.5pt\hbo time lag between the formation x{$\sim$}}\raise 1.5pt\hbox{$<$}\;}
\def\la{\mathrel{\hbox{\rlap{\hbox{\lower4pt\hbox{$\sim$}}}\hbox{$<$}}}}
\def\ga{\mathrel{\hbox{\rlap{\hbox{\lower4pt\hbox{$\sim$}}}\hbox{$>$}}}}

\begin{document}
\begin{frontmatter}
\title{Pre-main-sequence population in NGC 1893 region: X-ray properties}
\author[ari]{A. K. Pandey}
\cortext[cor3]{Corresponding author}
\author[lam]{M. R. Samal}
\author[ari]{Ram Kesh Yadav\corref{cor3}}
\ead{ramkesh@aries.res.in}
\author[nar]{Andrea Richichi}
\author[ari]{Sneh Lata}
\author[ari]{J. C. Pandey}
\author[tifr]{D. K. Ojha}
\author[tw]{W. P. Chen}
\address[ari]{Aryabhatta Research Institute of Observational Sciences (ARIES), Manora Peak, Nainital, India 263 002}
\address[lam]{Laboratoire d'Astrophysique de Marseille~-~LAM, Universit\'{e} d'Aix-Marseille $\&$ CNRS, UMR7326, 13388 Marseille Cedex 13, France}
\address[nar]{National Astronomical Research Institute of Thailand, 191 Siriphanich Bldg., Huay Kaew Rd., Suthep, Muang, Chiang Mai 50200, Thailand}
\address[tifr]{Tata Institute of Fundamental Research, Mumbai - 400 005, India }
\address[tw]{Institute of Astronomy, National Central University, Chung-Li 32054, Taiwan}


\linespread{2}

\begin{abstract}
Continuing the attempt to understand the properties of the stellar content in the young cluster NGC 1893 we have carried out a comprehensive multi-wavelength study of the region. The present study focuses on the X-ray properties of T-Tauri Stars (TTSs) in the NGC 1893 region.  We found a correlation between the X-ray luminosity, $L_X$, and the stellar mass (in the range 0.2$-$2.0 \msun) of TTSs in the NGC 1893 region, similar to those reported in some  other young clusters, however the value of the power-law slope obtained in the present study ($\sim$ 0.9) for NGC 1893 is smaller than those ($\sim$1.4 - 3.6) reported in the case of TMC, ONC, IC 348 and Chameleon star forming regions. However, the slope in the case of Class III sources (Weak line TTSs) is found to be comparable to that reported in the case of NGC 6611 ($\sim$ 1.1). It is found that the presence of circumstellar disks has no influence on the X-ray emission. The X-ray luminosity for both CTTSs and WTTSs is found to decrease systematically with  age (in the range $\sim $ 0.4 Myr - 5 Myr). The decrease of the X-ray luminosity of TTSs (slope $\sim$ -0.6) in the case of NGC 1893 seems to be faster than observed in the case of other star-forming regions (slope -0.2 to -0.5). There is indication that the sources having relatively large NIR excess have relatively lower $L_X$ values. TTSs in NGC 1893 do not follow the well established X-ray activity - rotation relation as in the case of main-sequence stars. 
\end{abstract}


\end{frontmatter}

\section{Introduction}

With the advent of new data from deep optical, near-infrared (NIR), mid-infrared (MIR) and X-ray surveys,  a considerable interest has been evolved to study low mass young stellar objects (YSOs) as well as to study the evolution of circumstellar disks around the YSOs associated with young star clusters (e.g., Prisinzano et al. 2011, 2012, Guarcello et al. 2012, Pandey et al. 2008, 2013, Haisch et al. 2001, Sicilia-Aguilar et al. 2006, Carpenter et al. 2006, Chauhan et al. 2009, 2011). Most of the disk evolution studies available in the literature are based on the NIR/ MIR excess which traces the evolution of inner accretion disks and it is found that in majority of the cases the inner disk does not last for more than $\sim$ 6 Myr (see e.g., Haish et al. 2001).

Most of the pre-main-sequence (PMS) stars are T-Tauri stars (TTSs) and have masses $\leq$ 3 M$_\odot$. The TTSs are further divided in to weakline TTSs (WTTSs) and classical TTSs (CTTSs). The H$\alpha$ and NIR excess signatures in CTTSs indicate the existence of a well-developed circumstellar disk actively interacting with the central star. Strong H$\alpha$ emission (EW $>$ 10 \AA) in CTTSs is attributed to the magnetospheric accretion of the innermost disk matter on to the central star (Edwards et al. 1994; Hartmann, Hewett \& Calvet 1994; Muzerolle, Calvet \& Hartmann 2001, and references therein). On the other hand, the weak H$\alpha$  emission (EW $<$ 10 \AA) in WTTSs, which lack disks (or, at least inner disks), is believed to originate from their chromospheric activity (e.g. Walter et al. 1988; Mart’in 1998).

Since both types of TTSs (CTTSs and WTTSs) are strong X-ray emitters, the X-ray observations of active star-forming regions and young Galactic clusters play an important role in the studies of the star formation process and the properties of young PMS stars (e.g., Feigelson \& Decampli 1981, Guarcello et al. 2012 and references therein). In fact, the level of X-ray emission in PMS stars, which is higher than that of field main-sequence (MS) stars, provides a very efficient mean of selecting stars associated with star-forming regions. However, the studies  based on H$\alpha$ emission may fail to detect part of the WTTS population, which can easily be identified in X-rays. 

Several X-ray studies of low-mass PMS stars have been carried out which revealed very strong X-ray activity, exceeding the solar levels by several orders of magnitude  (e.g., Feigelson \& Montmerle 1999; Feigelson et al. 2002; Favata \& Micela 2003; G\"{u}del 2004; Caramazza et al 2012). The influence of a circumstellar disk, and particularly the influence of accretion on X-ray activity has been of special interest. X-ray studies of PMS stars in star-forming regions have yielded contradictory results (see e.g., Stelzer \& Neuh\"{a}user 2001, Feigelson et al. 2002, Flaccomio et al. 2003, Preibisch et al. 2005, Telleschi et al. 2007, Guarcello at al. 2012). 

The TTSs show variation in their brightness in all wavelengths, from X-ray to infrared and have highly elevated levels of X-ray activity, however the relation between rotation and X-ray activity in TTSs remained unclear until recently. Stelzer \& Neuh\"auser (2001) have found that in the case of Taurus - Auriga region, Pleiades and Hyades, the rotation and X-ray emission are clearly correlated in the sense that faster rotators are more luminous. However, majority of the results do not find any correlation between rotation and X-ray activity (e.g., Flaccomio et al. 2003,  Rebull et al. 2006, Alexander \& Preibisch 2012).

A number of data sets dealing with the observational relationship between rotation and magnetic activity for young stellar clusters with ages as young as 30 Myr (e.g., Prosser et al. 1996; Stauffer et al. 1997; Jeffries et al. 2011) are available, however the data is still lacking for younger clusters having ages $\leq$ 5 Myr. In view of the contradictory results as discussed above regarding the effect of the disk on the X-ray luminosity of CTTSs and WTTSs as well as the lack of observational studies on the relationship between rotation and the X-ray properties, we feel that the young cluster NGC 1893 having an age of $\sim$ 2-4 Myr and located at a distance of 3.25 kpc (Sharma et al. 2007, Pandey et al. 2013, hereafter Pa13 ), is very well-suited in this respect. The cluster, located at the center of the Aur OB2 association, is associated with the \hii region IC 410. The cluster contains at least five O-type stars along with a rich population of PMS stars (see e.g. Pa13). Several photometric and low resolution spectroscopic studies of the region have been carried out (for details see Sharma et al. 2007, Prisinzano et al. 2011: hereafter P11) which manifest that star formation is still ongoing in the region and therefore NGC 1893 is an ideal object to study the  X-ray properties of the PMS stars.  

Caramazza et al. (2012) analyzed the X-ray properties of the Class II and Class III sources identified by P11 in the NGC 1893 region. They found that Class III stars appear intrinsically more X-ray luminous than Class II stars and concluded that this effect may be due to the presence of the  ”magnetically connected” disk itself or to the ongoing accretion from the disk to the star. However, in Pa13 we have shown that the sample of the Class II and Class III  YSOs in NGC 1893 region identified by P11 on the basis of  {\it Spitzer} MIR data was contaminated by the field stars. Chauhan et al. (2011) have also found a significant amount of contamination in the sample of YSOs  identified by Koenig et al. (2008) on the basis of {\it Spitzer} data in the W5 E \hii region. Pa13 and Chauhan et al. (2011) used NIR $(J - H)/(H - K)$ color - color (NIR-CC) diagram to avoid the contamination due to field star population in the YSO sample selected on the basis of {\it Spitzer} data. Hence it is considered worthwhile to re-analyze the X-ray properties of Class II and Class III sources and the relation between rotation and X-ray activity in TTSs  using the sample of the YSOs identified by Pa13 and the rotation period of the  YSOs determined by Lata et al. (2012, 2013).

\section {Data: Sample of YSOs}

On the basis of optical, NIR, MIR and X-ray data, P11 identified 1034 and 442 Class II and Class III YSOs, respectively. However, in our previous study (Pa13) we have shown that the sample of YSOs by P11 is significantly contaminated by the non-PMS stars. Pa13 have removed contamination in the sample of P11 using the NIR-CC  diagram. They found 367 and 246 Class II and Class III sources, respectively with optical counterparts. The $V/(V - I)$ color-magnitude diagram (CMD) of the 367 and 246 Class II and Class III sources (figure 14 of Pa13) reveals that the contamination due to field stars is greatly reduced, however still we noticed a few sources classified as Class II sources by P11 below the 5 Myr PMS isochrone. To further avoid the contamination due to background stars, we have considered only those sources as the PMS objects associated with NGC 1893 which are located above the 5 Myr isochrone (cf. figure 14 of Pa13). This selection yields 298 and 240 Class II and Class III sources, respectively. This sample is used to study the X-ray properties of the YSOs in the NGC 1893 region. 

NIR-CC diagram of Class II and Class III sources used in the present study is shown in Fig. 1, where the thin and thick dashed curves represent the unreddened MS and giant branches (Bessell $\&$ Brett 1988), respectively. The dotted line indicates the locus of intrinsic CTTSs (Meyer et al. 1997). The curves and the data are in the California Institute of Technology (CIT) photometric system. The parallel dashed lines are the reddening vectors drawn from the tip of the giant branch (left reddening line), from the base of the MS branch (middle reddening line) and from the base of the tip of the intrinsic CTTSs line (``right reddening line''). The extinction ratios $A_J/A_V = 0.265, A_H/A_V = 0.155$ and
 $A_K/A_V=0.090$ have been adopted from Cohen et al. (1981). Following Ojha et al. (2004) we classify NIR-CC diagram into three regions. The sources lying in the `F' region (located between the left and middle reddening lines) could be either field stars (MS stars, giants) or Class III and Class II sources with small NIR excesses. The `T' region, located between the  middle and right reddening lines' contains mostly CTTSs (Class II objects). There may be an overlap in NIR colors of Herbig Ae/Be stars and CTTSs in the `T' region (Hillenbrand et al. 1992). The `P' region is redward of the `T' region and the sources lying in this region are  likely Class I objects. The age and mass of each YSO have been estimated by Pa13 (cf. their table 1) using the $V, (V - I)$ CMD. The effect of random errors due to photometric errors and reddening estimation in determination of ages and masses were estimated by propagating the random errors to their observed estimation by assuming normal error distribution and using the Monte-Carlo simulations.
 

\begin{figure}
\centering
\resizebox{9cm}{9cm}{\includegraphics{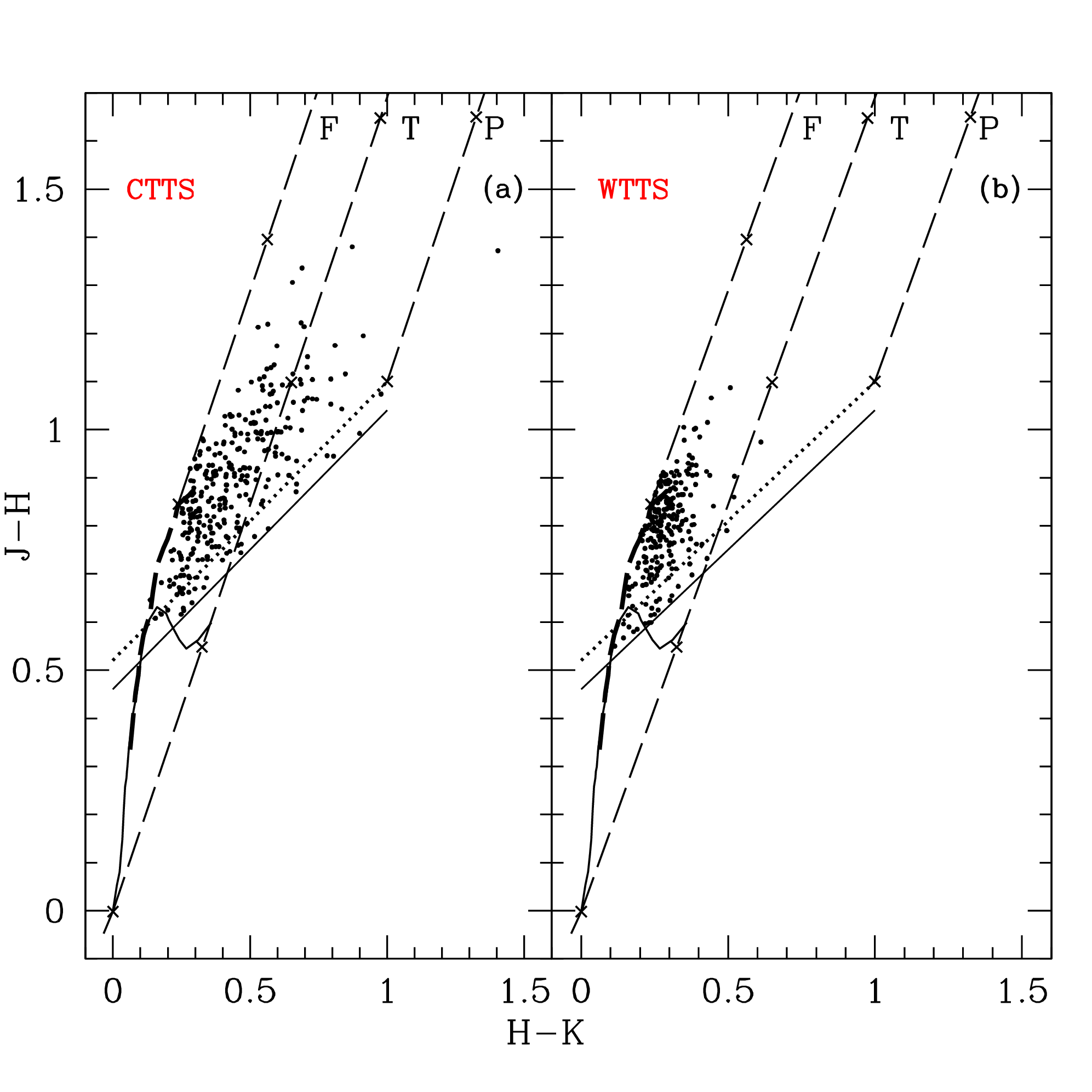}}

\caption {NIR-CC diagram for Class II (CTTSs) (left panel) and Class III (WTTSs) sources (right panel). The data have been taken from Pa13.  The locus for dwarfs (thin solid curve) and giants (thick dashed curve) are from Bessell \& Brett (1988). The dotted line indicates the locus of intrinsic CTTSs (Meyer et al. 1997). Dashed straight lines represent the reddening vectors (Cohen et al. 1981). The crosses on the dashed lines are separated by $A_V$ = 5 mag. The plots are classified into three regions, ‘F’, ‘T’ and ‘P’. The sources located in the ‘F’ region are likely to be the reddened field stars, WTTSs or CTTSs with little or no NIR excess. The sources in the ‘T’ region are considered to be candidate CTTSs with NIR excess and sources in the ‘P’ region are the candidate Class I objects. Considering the errors in the data, we consider only those sources as PMS sources associated with the region which lie above the continuous line (cf. Pa13).}
\end{figure}

The X-ray data has been taken from Caramazza et al. (2012). X-ray luminosity distribution of Class II (CTTSs) and Class III (WTTSs) sources used in the present study, for the  mass range 0.2$-$2.0 \msun, is shown in Fig. 2 which shows  a similar  distribution for both the samples. The bias due to low mass incompleteness was checked by plotting the distribution for the mass range 0.35$-$2.0 \msun and we again found the same distribution for both the samples indicating that the Class II and Class III sources have indistinguishable X-ray properties. Our previous study (Pa13) using the present sample has also shown that the CTTSs and WTTSs are coeval and have indistinguishable optical properties.


\begin{figure}
\centering
\resizebox{9cm}{9cm}{\includegraphics{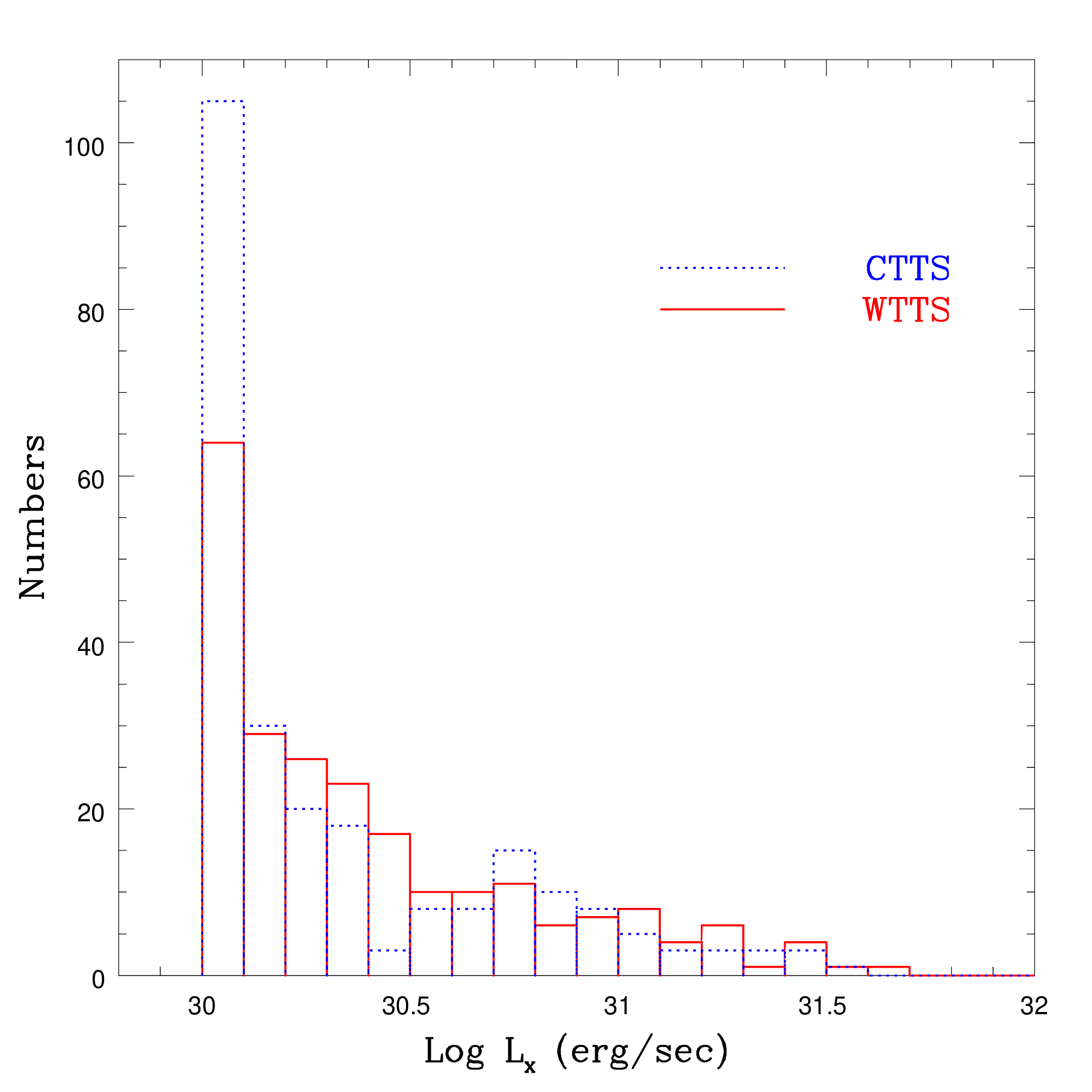}}

\caption {X-ray luminosity distribution of Class II (CTTSs) and Class III (WTTSs) sources used in the present study for the mass range 0.2$-$2.0 \msun. }
\end{figure}

\section {Results}
\subsection {Correlation with mass}

In Fig. 3 we plot the X-ray luminosity, $L_X$, as a function of the stellar mass. The CTTSs (Class II) and WTTSs (Class III) are plotted with filled and open circles, respectively. An increasing trend with a significant scatter in the $L_X$ with stellar mass  is apparent for the entire sample of the TTSs. A linear regression fit to the function log($L_X$) = $a +  b*log(M)$, in the mass range 0.2$-$2.0 \msun, yields $a$ = 30.71$\pm$ 0.04, 30.68 $\pm$ 0.06 and  30.74 $\pm$ 0.05 and $b$ = 1.16 $\pm$ 0.10, 1.14 $\pm$ 0.15 and 1.13 $\pm$ 0.13, respectively, for the whole sample of the TTSs, CTTSs and WTTSs. However, a careful inspection of Fig. 3 reveals that the lower envelope of the distribution  in the mass range 0.2 $\leq$ M/$M_\odot$ $\leq$ 0.35 is rather flat. The sources around lower envelope of 0.2 $\leq$ M/$M_\odot$ $\leq$ 0.35 have log($L_X$ $\simeq$ 29.7). This may indicate a bias in the sample in the sense that only X-ray luminous low mass sources might have been detected. A linear regression fit in the mass range 0.35$-$2.0 \msun yields $a$ = 30.69$\pm$ 0.05, 30.65 $\pm$ 0.08 and  30.72 $\pm$ 0.06 and $b$ = 1.05 $\pm$ 0.16, 0.99 $\pm$ 0.28 and 1.07 $\pm$ 0.20, respectively, for the whole sample of the TTSs, CTTSs and WTTSs. The results for two mass ranges are almost same within error.

\begin{figure}
\centering
\resizebox{9cm}{9cm}{\includegraphics{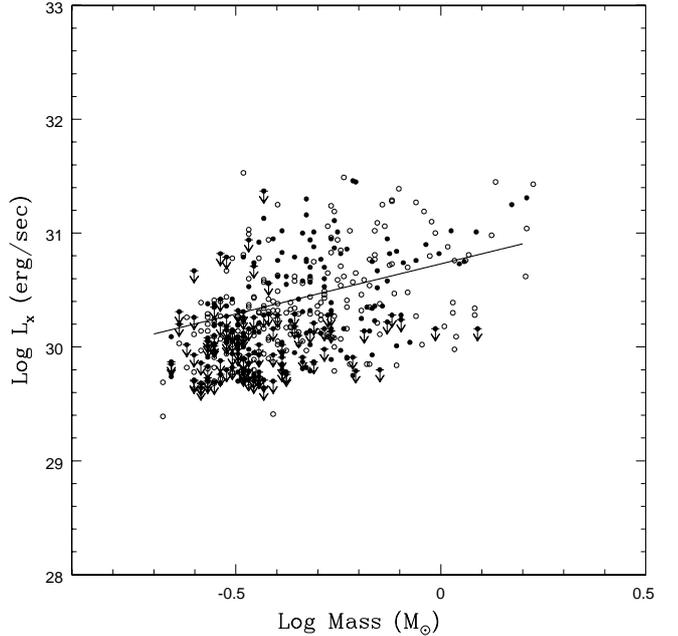}}

\caption{ Mass versus $L_X$ in logarithmic scale for CTTSs (filled circles), WTTSs (open circles). The downward-pointing arrows represent the sources with upper limit. The continuous line represents linear fit to the whole data obtained using the ASURV package.}
\label{NGC6823_field.ps}
\end{figure}

The sample of Class II sources (CTTSs) has a significant number of sources with upper limit. To account for upper limits of nondetections we used ASURV package as described by Preibisch and Feigelson (2005). The values of the slope $b$ in the mass range 0.2$-$2.0 \msun, comes out to be $b$ = 0.88$\pm$ 0.11, 0.51 $\pm$ 0.20 and 1.13 $\pm$ 0.13, respectively, whereas the values of $a$ are found to be 30.73$\pm$ 0.04, 30.71 $\pm$ 0.07 and  30.74 $\pm$ 0.05, respectively, for the whole sample of TTSs, CTTSs and WTTSs. The correltion is found to be $\sim 0.50$ with a probability of correlation greater than 99.9\% in all the cases.

In the case of Taurus molecular cloud (TMC), Telleschi et al. (2007) have found $a$ = 30.33 $\pm$ 0.06; 30.13 $\pm$ 0.09; 30.57 $\pm$ 0.09 and $b$ = 1.69 $\pm$ 0.11; 1.70 $\pm$ 0.20; 1.78 $\pm$ 0.17, respectively for the whole TTSs, CTTSs and  WTTSs. In the case of ONC, Preibisch et al. (2005) also found a linear regression for the whole TTSs sample (M $\leq$ 2\msun) with $a$ = 30.37 $\pm$ 0.06 and $b$ = 1.44 $\pm$ 0.10,  whereas Feigelson et al. (2003) reported $a$ = 30.33 and $b$ = 1.5 for the TTSs samples (M $\leq$ 3\msun) in the ONC.  The value of power-law slope $b$  for the TTSs in the Chamaeleon star-forming region in the mass range 0.6 - 2 \msun is reported to be  $b$ = 3.6 $\pm$ 0.6 (Feigelson et al. 1993) and in the very young stellar cluster IC 348, the slope $b$ in the mass range 0.1 - 2 \msun is found to be 1.97  $\pm$ 0.2 (Preibisch \& Zinnecker 2002). In the case of NGC 2264, the linear fit for the detected sources yields $a$ = 30.6 $\pm$ 0.3 and $b$ = 0.8 $\pm$ 0.1 (Dahm et al. 2007). However, taking into account the X-ray non-detections the slope is found to be steeper ($b$ = 1.5 $\pm$ 0.1).

The values of the power-law slopes reported in the case of TMC, ONC, IC 348 and Chamaeleon star-forming regions are higher than that obtained in the present study for NGC 1893. One possible reason for the difference could be  the bias in the sample, i.e. the present sample may have only X-ray luminous low mass stars, however we have shown above that the slope $b$ does not seem to be affected by the bias in the sample. Recently, Guarcello et al. (2012) for Class III stars  having mass M $\leq$  0.8\msun in NGC 6611 have found $a = 30.9 \pm 0.1$ and $b = 1.1 \pm 0.3$, whereas the distribution of stars more massive than 0.8 M$_\odot$ is flatter with a slope of 0.4 $\pm$ 0.2. The intercept of Class III sources at 1 M$_\odot$ (log$M/M_\odot$ = 0) has comparable value to that of Class II sources indicating that the presence of circumstellar disks has no influence on the X-ray emission. This result is in agreement with that reported by Feigelson et al. (2002) and is in contradiction with those reported by Stelzer \& Neuh\"{a}user (2001), Preibisch et al. (2005) and Telleschi et al. (2007). Caramazza et al. (2012) on the basis of X-ray luminosity functions (XLFs) have concluded that CTTSs in NGC 1893 are globally less X-ray active than the WTTSs. As we have already shown in our previous study (Pa13) that the sample of Class II sources used by P11 is strongly contaminated by field population, the XLF of Class II sources by Caramazza et al. (2012) may not be representing the true Class II XLF.

\subsection {Evolution of X-ray emission}


\begin{figure}
\centering
\resizebox{9cm}{9cm}{\includegraphics{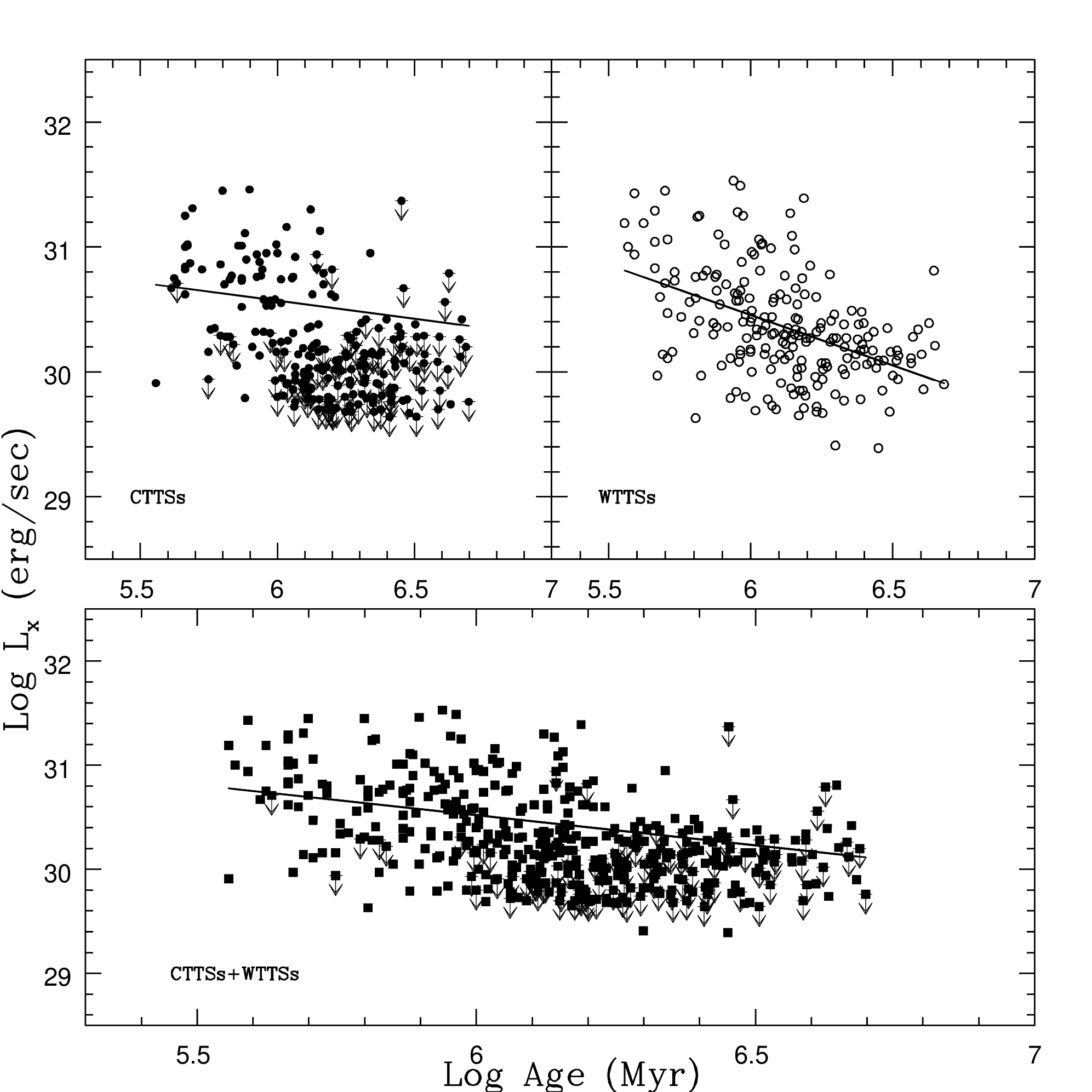}}

\caption {X-ray luminosity versus stellar age for CTTSs and  WTTSs  having masses in the range 0.2-2.0 $\msun$. The downward-pointing arrows represent the sources with upper limit. The lines represent linear fit to the data obtained using the the ASURV package.}
\label{NGC6823_field.ps}
\end{figure}

The age distribution of identified YSOs in NGC 1893 reveals a non-coeval star formation in the cluster (e.g. Pa13). Keeping the caveats in mind  (e.g. age determination is model dependent, systematic errors due to presence of unresolved binaries etc.) as discussed by Preibisch \& Feigelson (2005) as well as probable bias in the sample as discussed in Sec 3.1, we present evolution of X-ray emission in CTTSs and WTTSs in the mass range $0.2 \le M/M_\odot \le 2.0$ in Fig. 4.  Although the X-ray emission as discussed above, is correlated with mass, we could not study the evolution of X-ray emission for mass stratified sample because of limited data sample. The X-ray luminosity for both CTTSs and WTTSs 
can be seen to decrease systematically with  age (in the age range $\sim $ 0.4 to 5 Myr). A linear regression fit to the function log$(L_X)$ = a +  b*log$(age)$ yields $a$ = 35.26 $\pm$ 0.45, 35.06 $\pm$ 0.63 and  35.29 $\pm$ 0.65; and $b$ = -0.81 $\pm$ 0.07, -0.78 $\pm$ 0.10 and -0.81 $\pm$ 0.11, respectively for the whole sample of the TTSs, CTTSs and WTTSs. All the slopes are same within the errors. We have also computed linear regression using the ASURV package and found $a$ = 34.00 $\pm$ 0.50, 32.31 $\pm$ 0.81 and  35.29 $\pm$ 0.65; and $b$ = -0.58 $\pm$ 0.08, -0.29 $\pm$ 0.13 and -0.81 $\pm$ 0.11, respectively, for the whole sample of the TTSs, CTTSs and WTTSs. 

The slopes obtained in the case of NGC 1893  are slightly steeper in comparison to those (-0.2 to - 0.5)reported by Preibisch \& Feigelson (2005) and Telleschi et al. (-0.36 $\pm$ 011; 2007). This difference in the slope could be more significant in the case of an unbiased sample, where we expect more low luminous low mass X-ray sources, which would give more higher value for the slope. If X-ray luminosities of accreting PMS stars are systematically lower than non-accreting PMS stars (e.g. Flaccomio et al. 2003; Preibisch et al. 2005) and if CTTSs evolve to the WTTSs, one might expect that in the case of CTTSs $L_X$ should increase with age rather than decrease. However, the evolution of CTTSs up to $\sim $ 5 Myr (Fig. 4) does not show any sign of increase in X-ray luminosity. This result seems to contradict the notion that CTTSs evolve to WTTSs.

\subsection {Accretion disk and X-ray luminosity}


\begin{figure}
\centering
\resizebox{9cm}{9cm}{\includegraphics{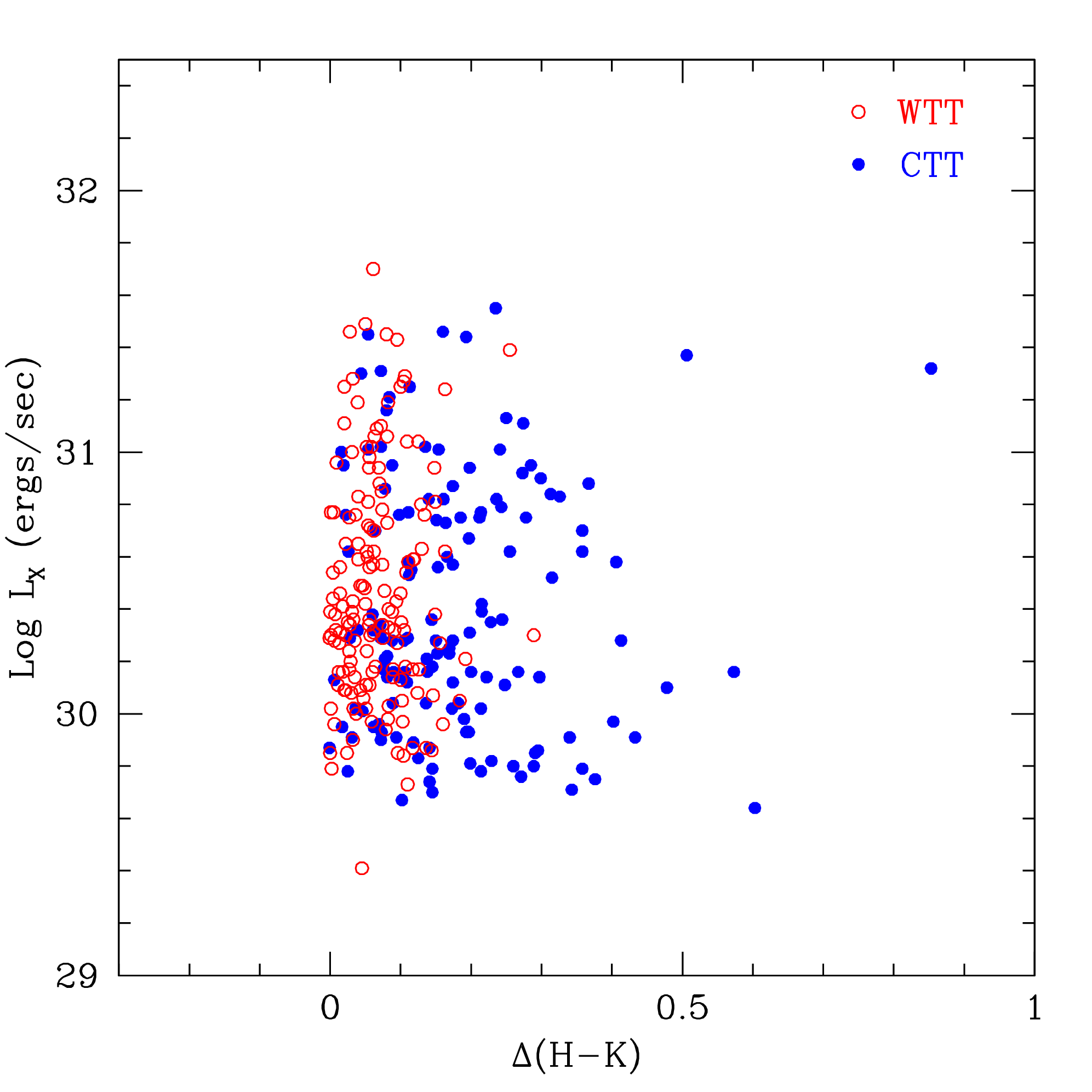}}
\caption { NIR excess $\Delta (H-K)$ versus X-ray luminosity of Class II (CTTSs) and Class III (WTTSs) sources in the mass range 0.2-2.0 \msun. }
\label{NGC6823_field.ps}
\end{figure}

Rebull et al. (2006) have pointed out that two processes associated with disk accretion can affect the observed X-ray properties of solar-type PMS stars. In the first case local heating of the stellar photosphere at the "footprints" of magnetospheric funnel flows, where highly supersonic  ($v$ $>$ 200 km s$^{-1}$) gas lands on the stellar surface  (e.g., Kastner et al. 2002; Schmitt et al. 2005), thus producing shocks and local heating, and as a result produces soft X-rays. This process would qualitatively produce an increase in X-ray emission above coronal activity levels as disk accretion rates increase. In the second case absorption of keV coronal X-ray emission by gas associated with accretion-driven outflows (Walter \& Kuhi 1981) would produce a decrease of X-ray emission with increasing accretion rate, since the column density, and hence the X-ray  optical depth of out-flowing material, is expected to increase with increasing accretion rate. 

There has been several attempts to study the relationships between circumstellar disks, disk accretion, and X-ray emission, however the results are contradictory (see e.g. Rebull et al. 2006 and references therein).  Kastner et al. (2002) and Schmitt et al. (2005) found evidence of enhanced X-ray emission associated with transient increases in disk accretion rate. This fact indicates that at least some X-ray emission may be associated with accretion as opposed to coronal activity. Majority of the results, however, available in the literature suggest that there is no evidence of excess X-ray emission for accreting PMS stars. Rather, there  seems to be anti-correlation between the presence of an accretion disk and total X-ray luminosity (e.g., Rebull at al. 2006).

Walter \& Kuhi (1981) reported an inverse correlation between soft X-ray flux and H$\alpha$ emission for CTTSs. Assuming that the observed equivalent width of H$\alpha$ serves as a surrogate for mass outflow rate, Walter \& Kuhi (1981) argued that the apparent decrease of X-ray emission with increasing H$\alpha$ equivalent width reflected the effects of X-ray absorption by outflowing gas. Flaccomio et al. (2003) also found that a larger fraction of stars with active accretion signatures have fainter $L_X$ values than non-accretors. However, Dahm et al. (2007), in the case of NGC 2264, do not find any correlation between  H$\alpha$ equivalent width and X-ray luminosity of TTSs. The NIR excess (e.g. (H-K) excess) is a useful indicator for the presence of disk, hence can be used to look for correlations of X-ray emission with the presence/absence of accretion disks for which there is no direct evidence of accretion. On the basis of NIR excess stars, Feigelson et al. (2002, 2003), Rebull et al. (2006), Dahm et al. (2007) have concluded that the presence or absence of infrared excess, has no discernible effect on the distribution of X-ray luminosities. 

To study the effect of disk on the X-ray emission we define NIR excess $\Delta (H-K)$ as the horizontal displacement from the left reddening vector of `F' region in the NIR-CC diagram as described in Fig. 1. In Fig. 5 we plot $L_X$ of CTTSs (filled circles) and WTTSs (open circles) as a function of NIR excess $\Delta (H-K)$.  Although the scatter is large, however Fig. 5 manifests that CTTSs having relatively larger $\Delta (H-K)  \gtrsim$ 0.4 - 0.5 mag have smaller value  of log $L_X$ ($\sim$ 30), whereas the CTTSs having $\Delta (H-K)$ $\le$  0.1 mag have log $L_X$  values up to $\sim$ 31.5. It is interesting to point out that the upper bound for the Class II sources (CTTSs) for $\Delta (H-K)  \gtrsim 0.2$ appears to decrease with increasing value of $\Delta (H-K)$. A similar trend has been noticed by Rebull at al. (2006) in the $(I-K)$ excess versus log $L_X$ plot of NGC 2264. This may be indicative of that the sources having relatively large NIR excess have relatively lower $L_X$ values. However, Feigelson et al. (2003) on the basis of  $\Delta (I-K)$ excess in the case of Orion nebula cluster do not find any correlation between X-ray emission and the presence of disk. The distribution of WTTSs in Fig. 5 does not show any obvious correlation. Since $U$-band data is not available for these TTSs, it is not possible to estimate the mass accretion rate for the CTTSs studied here. Sicilia-Aguilar et al. (2005) have presented NIR data and mass accretion rate for a few TTSs of young cluster Tr 37. The data reveals a positive correlation between NIR excess $\Delta (H-K)$ and mass accretion rate. Assuming that the same correlation exists in the case of NGC 1893, we can presume that the mass accretion rates are inversely correlated with X-ray luminosity.

\subsection {Rotation and X-ray activity relation}

Both WTTSs and CTTSs show variation in their brightness. These variations are found to occur at all wavelengths, from X-ray to infrared. The variability time-scale of TTSs ranges from a few minutes to years (Appenzeller \& Mundt 1989). Several mechanisms such as circumstellar gas and dust (remnant of parent molecular cloud), accretion and magnetic field have been suggested to explain the photometric variations (Herbst et al. 1994). The variations in the brightness of TTSs are most  probably due to the presence of spots (cool or hot) on stellar surface and circumstellar disk. The cool spots on the surface of the stars are  produced by the emergence of stellar magnetic fields on the photosphere, and are thus indicators of magnetic activity. The cool spots on the photosphere rotate with stars hence are responsible for brightness variation in WTTSs. These objects are found to be fast rotators as they have either thin or no circumstellar disk.  The hot spots on the surface of young stars are the consequence of accretion process (Lynden-Bell \& Pringle 1974; Koenigl 1991; Shu et al. 1994). Irregular or non-periodic variations are produced because of changes in the accretion rate. The time-scales of varying brightness range from hours to years. The hotspots cover a smaller fraction of the stellar surface but higher temperature causes larger brightness variations (Carpenter, Hillenbrand \& Skrutskie 2001). The accreting CTTSs show a complex behaviour in their optical and NIR light curves (Scholz et al. 2009).

Although the YSOs in all evolutionary stages, from Class I protostars to TTSs, have highly ($\sim$ 10$^3$-10$^4$ times) elevated levels of X-ray activity, the physical origin of this X-ray activity remains poorly understood. However, there is strong evidence that in most TTSs the X-ray emission originates from magnetically confined coronal plasma (e.g., Preibisch et al. 2005), it is still not clear about the dynamo processes that create the required magnetic fields. It is well established that the level of the magnetic activity, i.e., the strength of the X-ray emission in MS stars is mainly determined by their rotation rate (e.g., G\"udel, Guinan, \& Skinner 1997; Gaidos 1998; Pallavicini et al. 1981; Pizzolato et al. 2003). The relation between rotation and X-ray activity in TTSs remained unclear until recently, since in most studies of star-forming regions the number of X-ray detected TTSs with known rotation periods was too small to draw any conclusion. The Chandra Orion Ultradeep Project (COUP, Getman et al. 2005) and the XMM-Newton Extended Survey of  the Taurus Molecular Cloud (XEST, G\"udel et al. 2007) provided very sensitive X-ray data sets for large samples of TTSs. These data revealed that the TTSs in Orion as well as in Taurus do not follow the relation between rotation period and X-ray luminosity for MS  stars (Preibisch et al. 2005; Briggs et al. 2007). In addition the TTSs spin up during the first $\sim$ 10 - 30 Myr (Herbst \& Mundt 2005) does not lead to an increase in the X-ray luminosity. This places doubt on the solar-like dynamo activity scenario for TTSs (Alexander \& Preibisch, 2012).

In Fig. 6 we plot X-ray luminosity L$_X$ as a function of rotation period. The data of rotation period for 38 and 10  TTSs have been taken from Lata et al. (2012) and Lata et al. (2013, in preparation), respectively. The plot manifests that the TTSs in NGC 1893 do not follow the well established X-ray activity - rotation relation as in the case of main-sequence stars, i.e. increase in X-ray activity with decreasing rotation period. The least-squares fit to the data points in Fig. 6 yields a slope of $-0.06\pm0.09$ which indicates that the X-ray activity for the TTSs in NGC 1893 seems to be independent of the rotation period. 

Flaccomio et al. (2003) found for the first time that TTSs in ONC having rotation period show higher X-ray luminosities than TTSs do not have known rotational periods. The same has been found in the case of NGC 2264 (Rebull et al. 2006). Fig. 7 shows X-ray luminosity distribution of TTSs having rotation periods as well as TTSs without data for rotation periods. The TTSs with rotation periods show a uniform distribution for the X-ray luminosity whereas remaining TTS sample peaks at Log $L_X \sim$ 30.


\begin{figure}
\centering
\resizebox{9cm}{9cm}{\includegraphics{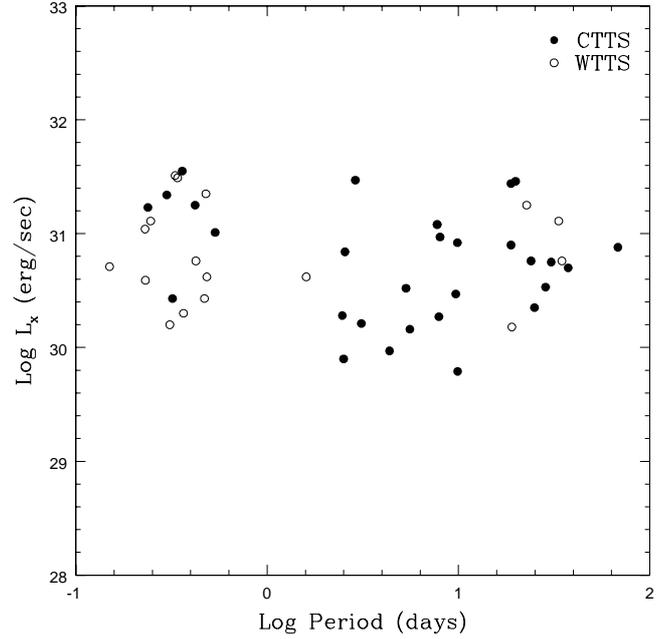}}

\caption {Plot of X-ray luminosity versus rotation period for TTSs having masses in the range 0.2-2.0 $\msun$. }
\label{NGC6823_field.ps}
\end{figure}


\begin{figure}
\centering
\resizebox{9cm}{9cm}{\includegraphics{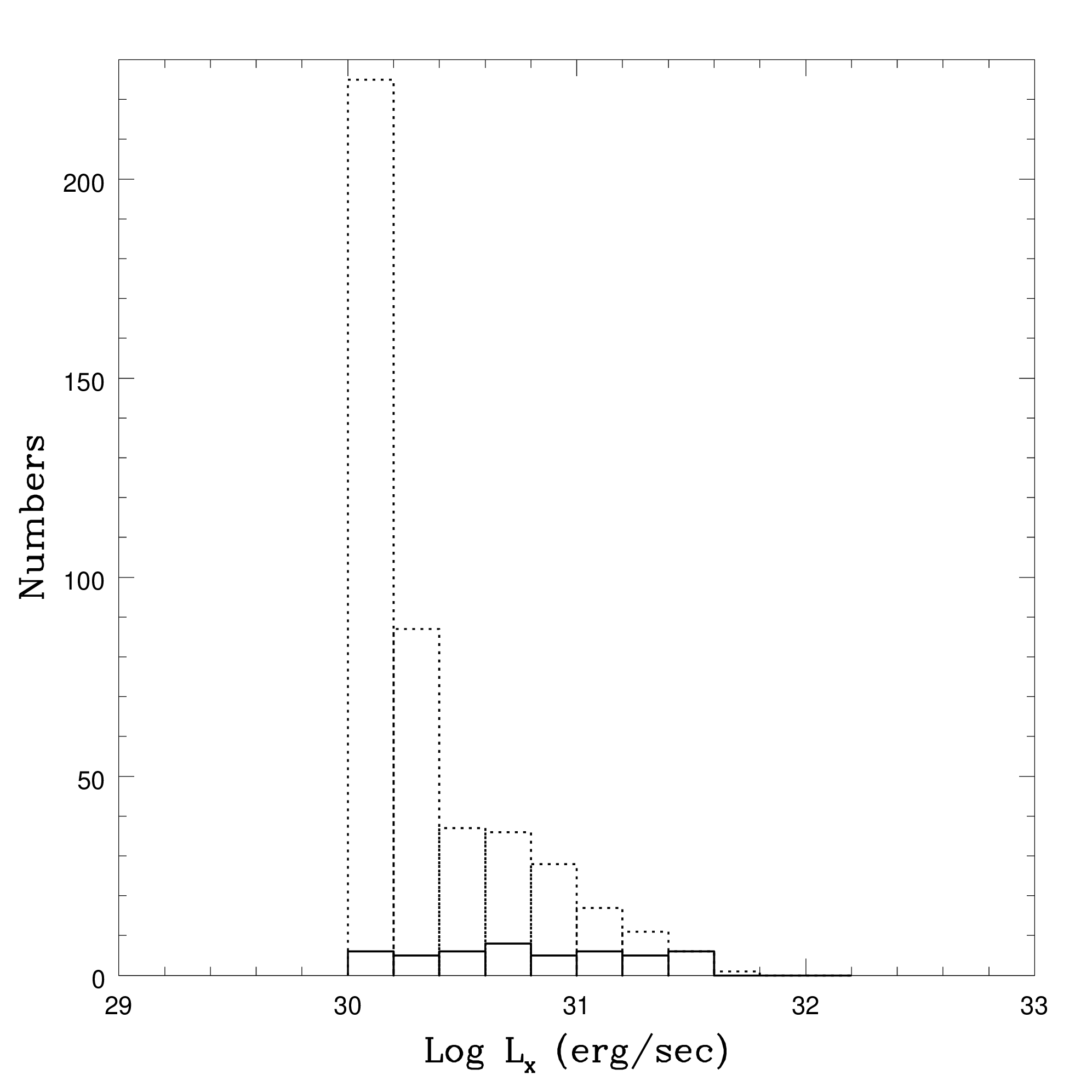}}

\caption {X-ray luminosity distribution of TTSs having known rotation periods (continuous line) as well as TTSs without data for rotation periods (dotted line). }
\label{NGC6823_field.ps}
\end{figure}

\section{Conclusions}

On the basis of a comprehensive multi-wavelength study of the star forming region NGC 1893, we continued our attempt to understand the properties of the stellar content in the young cluster NGC 1893. This paper is focused on understanding the X-ray properties of TTSs in the NGC 1893 region. The main results of the present work are as follow:

1) We found a correlation between X-ray luminosity, $L_X$, and the stellar mass (in the range 0.2$-$2.0 \msun)  of TTSs  of NGC 1893, as already has been reported by several authors, however the values of the power-law slopes ($\sim$ 3.6 - 1.4) reported in the case of TMC, ONC, IC 348 and Chameleon star forming regions are higher than that obtained in the present study ($\sim 0.5 - 1.1$) for NGC 1893. The slope ($\sim 1.1$) of the distribution of Class III sources in the case of NGC 1893 is found to be comparable to that reported in the case of NGC 6611. 

2) The intercept of Class III sources at 1 \msun (log$M/M_\odot$ = 0) in the present sample  has comparable value to that of Class II sources indicating that the presence of circumstellar disks has no influence on the X-ray emission. This result is in agreement with that reported e.g., by Feigelson et al. (2002) and is in contradiction with those reported by Stelzer \& Neuh\"auser (2001), Preibisch et al. (2005) and Telleschi et al. (2007). On the basis of XLFs Caramazza et al. (2012) have concluded that CTTSs in NGC 1893 are globally less X-ray active than the WTTSs. However, we have to keep in mind that Pa13 have already shown that the sample of Class II sources  by  Caramazza et al. (2012) is strongly contaminated by field population. 

3) The X-ray luminosity for both CTTSs and WTTSs is found to decrease systematically with  age (in the range $\sim $ 0.4 Myr - 5 Myr). The decrease of the X-ray luminosity (slope $\sim$ -0.6) in the case of NGC 1893 is faster than observed in the case other star forming regions (e.g., ONC: slope -0.2 - 0.5, Preibisch \& Feigelson 2005; TMC: $-0.36$ $\pm$ $0.11$, Telleschi et al. 2007).

4) It is found that the upper bound of the distribution of $\Delta (H-K)$ vs log $L_X$ for the Class II sources (CTTSs) having  $\Delta (H-K)$ $\gtrsim$ $0.2$  appears to decrease with increasing value of $\Delta (H-K)$. A similar trend has been noticed by Rebull at al. (2006) in the (I-K) excess versus log $L_X$ plot of NGC 2264. This may be indicative of that the sources having relatively large NIR excess have relatively lower $L_X$ values.  The X-ray properties of the Class II and Class III sources of the present sample are rather same because a significant number of Class II sources have weak NIR excess (cf. Fig. 2). Present result suggests that CTTSs having higher NIR excess (an indication of the presence of inner disk) could have lower X-ray luminosity because of higher  extinction due to X-ray absorption by circumstellar disks.

5) TTSs in NGC 1893 like TTSs in other young clusters (e.g., in ONC: Flaccomio et al. 2003, NGC 2264: Rebull et al. 2006, IC 348: Alexander \& Preibisch 2012) do not follow the well established X-ray activity - rotation relation as in the case of main-sequence stars. Present result is in contrast with Stelzer \& Neuha\..user (2001; Taurus - Auriga region, Pleiades and Hyades region) and Preibisch et al. (2005; Orion region) where an anti-correlation and correlation, respectively, was reported between the X-ray luminosity and rotation period.

\section*{Acknowledgments}
AKP is thankful to Prof. B. Soonthornthum Director, NARIT for encouraging the collaboration between ARIES (India) and NARIT (Thailand). This publication makes use of the data products from the 2MASS, which is a joint project of the University of Massachusetts and the Infrared Processing and Analysis Center/California Institute of Technology.

\end{document}